\documentclass[aip,apl,preprint]{revtex4-2}

\usepackage{amsmath}
\usepackage{graphicx}% Include figure files
\usepackage{dcolumn}% Align table columns on decimal point
\usepackage{bm}% bold math
\usepackage{hyperref}% add hypertext capabilities
\usepackage[T1]{fontenc}
\usepackage{textcomp}
\usepackage{xfrac}
\usepackage{comment}
\usepackage{xcolor}
\usepackage{siunitx}
\begin{document}

\preprint{}

\title{Frequency-Division Multiplexing in Magnonic Waveguides}% Force line breaks with \\
%%\title{Manuscript Title:\\with Forced Linebreak}% Force line breaks with \\
%%\thanks{A footnote to the article title}%

\author{Alina-Cristina Bunea}
\email{alina.bunea@imt.ro}
 \affiliation{National Institute of Research and Development in Microtechnologies - IMT Bucharest, Romania}%Lines break automatically or can be forced with \\

\author{Laurentiu Stoleriu}%
\affiliation{Department of Physics, Alexandru Ioan Cuza University, 700506 Iasi, Romania}%

\author{Giacomo Talmelli}%
\affiliation{Imec, 3001 Leuven, Belgium}

\author{Dan Neculoiu}%
 %\email{dan.neculoiu@imt.ro}
\affiliation{National Institute of Research and Development in Microtechnologies - IMT Bucharest, Romania}
\affiliation{National University of Science and Technology POLITEHNICA Bucharest}

\author{Christoph Adelmann}
\affiliation{Imec, 3001 Leuven, Belgium}

\author{Florin Ciubotaru}
\email{Florin.Ciubotaru@imec.be}
\affiliation{Imec, 3001 Leuven, Belgium}

\date{\today}% It is always \today, today,

\begin{abstract}
Frequency-division multiplexing is a key functionality for wave-based information processing, enabling multiple information channels to coexist within the same physical medium. Here, we experimentally investigate spin-wave multiplexing in a CoFeB waveguide using all-electrical excitation and detection. Two independently generated microwave signals are simultaneously coupled into the same spin-wave waveguide through inductive antennas and characterized using broadband vector network analyzer measurements. The transmission spectra obtained under single-channel and multiplexed operation exhibit excellent agreement, demonstrating that spin waves with different frequencies and wavelengths propagate simultaneously without measurable interaction in the linear regime. The observation is confirmed using both dual-sweep and sweep-plus-single-tone excitation schemes. A theoretical analysis based on linear superposition and phase-sensitive detection explains the absence of observable inter-channel interference for independent microwave sources. Micromagnetic simulations further confirm that the amplitudes and wavevectors of the individual spin-wave modes remain unchanged during co-propagation, demonstrating the absence of interaction over the entire propagation distance. The results provide direct experimental evidence that independent spin-wave channels can coexist in a single waveguide and support the implementation of frequency-division multiplexing in future magnonic computing and microwave signal-processing architectures. 

\end{abstract}

\maketitle

%\tableofcontents

Spin waves have emerged as promising information carriers for beyond-CMOS computing and microwave signal processing owing to their wave nature, nanoscale wavelengths, and potential for low-power operation. Their phase, amplitude, and frequency can be exploited to encode and process information, enabling functionalities that are difficult to achieve with conventional charge-based electronics \cite{serga2010,lan2015, pirro2021}. In particular, interference-based devices such as majority gates and logic networks have attracted considerable attention as building blocks for future magnonic circuits. \cite{khitun2010,talmelli2020,mahmoud2021,chumak2015, levchenko2026,kruglyak2010} Beyond logic applications, spin waves are also being explored for analog signal processing, neuromorphic computing, and microwave information technologies, where the coexistence of multiple wave states offers opportunities for highly parallel operation.\cite{chumak2015,pirro2021}

An important advantage of wave-based information processing is the possibility of frequency-division multiplexing (FDM), whereby several independent information channels coexist within the same physical medium. Such multiplexing techniques form the basis of modern communication systems and could significantly enhance the computational throughput and functionality of future magnonic hardware. In the linear regime, spin waves obey the superposition principle and are expected to propagate independently without mutual interaction, allowing multiple frequency channels to share the same waveguide. \cite{sarker2022} This capability would enable parallel data transmission and processing without increasing device footprint, making it particularly attractive for dense magnonic interconnects and large-scale spin-wave circuits.\cite{talmelli2020},\cite{mahmoud2021} Several magnonic multiplexing and demultiplexing concepts have been proposed and experimentally demonstrated using anisotropic spin-wave propagation, engineered magnetic-field landscapes, directional couplers, magnonic crystals, and frequency-selective routing structures. \cite{vogt2014, heussner2020,zhang2019,nikolaev2024,morozova2025,davies2015,davidkova2025,morozova2024} In particular, Zhang \textit{et al.} demonstrated frequency-division multiplexing in an inhomogeneously biased YIG waveguide, where spin waves of different frequencies propagated simultaneously in spatially separated channels created by a lateral magnetic-field gradient \cite{zhang2019}. Similarly, frequency-selective magnonic routing and demultiplexing have been demonstrated through the exploitation of anisotropic spin-wave dispersion and engineered magnetic environments, establishing the feasibility of parallel magnonic signal transport \cite{heussner2020,nikolaev2024,morozova2024}. However, in most of these studies, the individual frequency channels are either spatially separated or investigated using optical detection techniques such as microfocused Brillouin light scattering (\textmu-BLS).

\begin{figure*}
\includegraphics[width=1\textwidth]{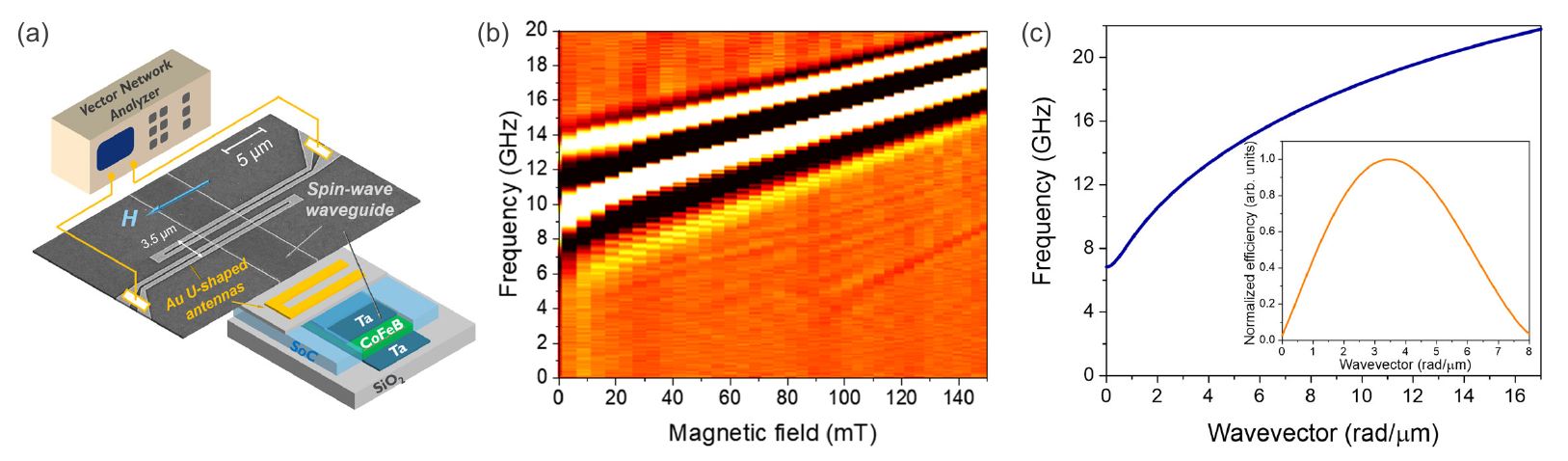}% Here is how to import EPS art
\caption{\label{fig1}(a) Microscope image of the investigated device and a schematic of the material stack together with the experimental setup based on a Vector Network Analyzer (VNA). (b) Bias-field derivative of the imaginary component of the S$_{21}$ microwave transmission parameter associated with spin-wave propagation. (c) Spin wave dispersion relation analytically calculated for an applied magnetic field $\mu_{0}$H = 43 mT and the measured material parameters. Inset: excitation efficiency of the U-shaped antenna as a function of the wavevector.}
\end{figure*}

Despite this progress, experimental investigations of spin-wave multiplexing using fully electrical excitation and detection schemes remain limited. In particular, the simultaneous propagation of spin waves generated by independent microwave sources within the same magnetic waveguide, and the extent to which different frequency channels remain noninteracting during propagation, have received relatively little experimental attention. In this work, we investigate spin-wave frequency multiplexing in a CoFeB waveguide using inductive antenna transducers and broadband microwave measurements. Unlike previous demonstrations relying on spatially separated channels or frequency-selective routing \cite{heussner2020,zhang2019,nikolaev2024,morozova2024,morozova2025}, the present study focuses on the simultaneous excitation and propagation of multiple frequency channels within the same magnetic conduit and their direct electrical characterization. Using two independently generated microwave excitation signals, we demonstrate that spin waves with different frequencies and wavelengths can coexist and propagate simultaneously without measurable interaction when operating in the linear regime. Micromagnetic simulations further confirm that this behavior is maintained over the entire propagation distance of the waveguide and is independent of the detection position. These results provide direct experimental and numerical evidence for the coexistence of multiple spin-wave frequency channels within a single waveguide, establishing frequency-division multiplexing as a viable approach for parallel magnonic computing, signal-processing, and communication architectures.

The studied device consists of a Ta/Co$_{40}$Fe$_{40}$B$_{20}$/Ta (3/30/3 nm) waveguide deposited on Si/SiO2 substrate by physical vapor deposition and patterned using ion-milling. A saturation magnetization of 1.36 MA/m was determined by vibrating-sample magnetometry and ferromagnetic resonance (FMR) experiments. The spin waves are generated and detected by two 500 nm wide Au U-shaped antennas patterned on top of the waveguide. The U-shaped antennas are connected to the ground and signal lines of a coplanar waveguide designed for a 50 $\Omega$ impedance \cite{talmelli2020}. An external magnetic field is applied transverse to the magnonic waveguide, as indicated in the Fig. 1(a).

In a first step, the spin-wave transmission characteristic was measured as a function of the external bias magnetic field using a vector network analyzer with an RF output power of 0.5 mW. The reflected (S$_{11}$) and transmitted (S$_{21}$) signals were measured as functions of both the applied magnetic bias field and frequency. Figure 1(b) presents the field-derivative of the imaginary part of the S$_{21}$ scattering parameter. The data indicates that the spin waves are generated in relatively wide bands, ~5 to ~7 GHz (at moderate magnetic fields), in agreement with the dispersion relations calculated analytically \cite{slavin} considering the experimental material parameters, the effective internal field determined from micromagnetic simulations, and the excitation efficiency of the U-shaped antenna. 

\begin{figure}
\includegraphics[width=0.5\textwidth]{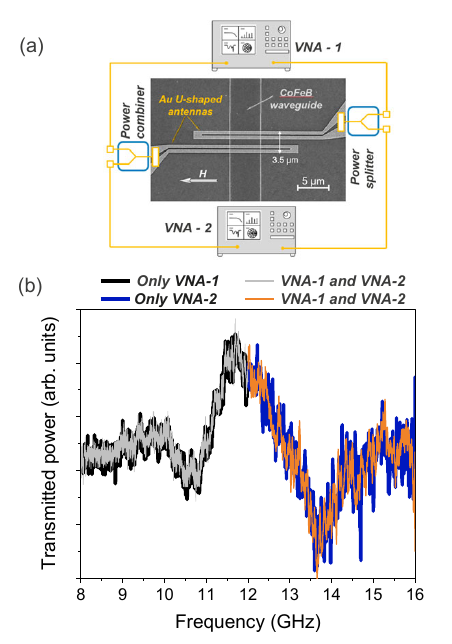}% Here is how to import EPS art
\caption{\label{fig2}(a) Microscope image of the investigated device and a schematic of the measurement setup consisting of two VNAs, power splitter and combiner. (b) Transmitted spin-wave power measured by the two VNAs that are either operating individually or simultaneously sweeping the frequency. VNA-1 is operating within the 8-12 GHz band, whereas VNA-2 is sweeping the frequency in the 12-16 GHz range. The applied magnetic field is $\mu_{0}$H = 43 mT.}
\end{figure}

In a frequency-division multiplexing experiment, two or more spin waves at different frequencies propagate simultaneously within the same magnetic conduit. To demonstrate that spin waves can coexist without interaction in the linear regime, we employed two vector network analyzers (VNAs) connected to the same device and operating in separate frequency bands (see Fig.~\ref{fig2}(a)). The output signals from both VNAs are combined using a power combiner and fed into the input antenna of the spin-wave device. At the output, the transmitted spin-wave signal is detected by a second U-shaped antenna, split by a power divider, and routed back to the two VNAs.
Both VNAs are triggered simultaneously and perform frequency sweeps with identical frequency steps (1 MHz) and output power (0.5 mW). The low excitation power was chosen to ensure operation within the linear regime and to avoid nonlinear spin-wave effects. VNA-1 sweeps the frequency range from 8 to 12 GHz, while VNA-2 simultaneously sweeps from 12 to 16 GHz. These frequency intervals were selected such that each VNA sweeps a part of the spin-wave transmission band at an applied magnetic field of 43 mT. As a result, two independent signals propagate through the device at all times, forming a frequency-multiplexed configuration.

Initially,  only one VNA was active and sweeping, while the other was set to an “RF off hold” state (no signal emitted). The measured data in Fig.~\ref{fig2}(b) shows the transmitted spin-wave signal over the full combined frequency range, plotted on a common frequency axis. The traces represent the difference in the magnitude of the transmission parameter S$_{21}$ measured at an applied magnetic field of 43 mT and at 0 mT (to remove the direct electromagnetic coupling between the two antennas). It should be noted that the two VNAs exhibit different noise floors, which results in a higher noise level in the 12–16 GHz range.

Subsequently, both VNAs were operated simultaneously, sweeping their respective frequency ranges under a common trigger. When multiple microwave sources are connected to the excitation antenna, the measured response depends on their mutual phase relationship. For phase-locked sources, the individual contributions add coherently, resulting in a vector addition of the corresponding complex transmission coefficients. In contrast, free-running sources exhibit a continuously varying relative phase. Since a VNA performs phase-sensitive detection referenced to its internal source, signal components generated by non-synchronized sources average out over the measurement time and contribute primarily to the measured power background and noise floor. Consequently, in the absence of nonlinear interactions, the transmission spectrum measured by a VNA remains dominated by the spin waves excited by its own source, while simultaneously propagating spin waves generated at different frequencies by independent microwave sources produce negligible changes in the measured complex response. This behavior constitutes the basis for frequency-division multiplexing in magnetic waveguides, where multiple spin-wave channels can coexist and propagate simultaneously without mutual interference.

The nearly identical transmission characteristics (within the noise limit) observed under single-source and dual-source operation indicate the absence of interaction between spin waves of different frequencies and wavelengths in the linear regime. This provides direct experimental evidence for frequency-division multiplexing of spin waves within a single waveguide, enabling simultaneous transmission without mutual interference.
In the linear excitation regime, the dynamic magnetization response follows the superposition principle. Consequently, spin waves at different frequencies propagate independently within the same magnetic waveguide, and the detected signal corresponds to the sum of the individual contributions. Therefore, the real and imaginary parts of  S$_{21}$ carry information about both the amplitude and the phase accumulated during propagation.

\begin{figure*}
\includegraphics[width=1\textwidth]{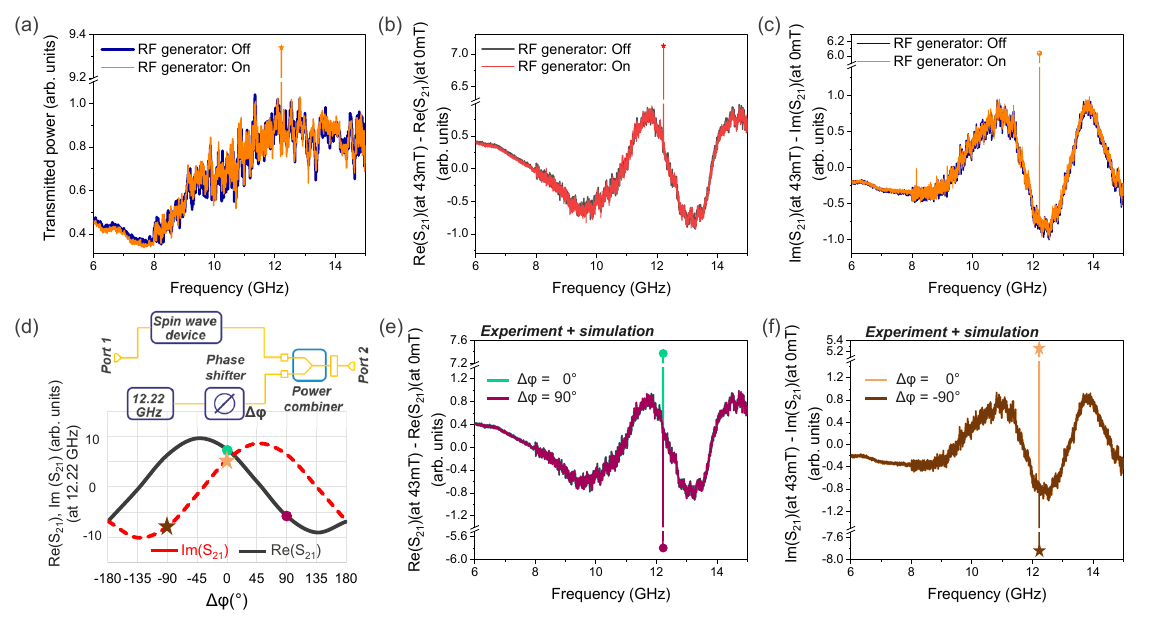}% Here is how to import EPS art
\caption{\label{fig3}(a) Measured spin-wave transmission power recorded by VNA-1 with the 12.22 GHz RF generator signal switched on and off. (b,c) Real and imaginary parts of the transmission parameter S$_{21}$ measured with and without the 12.22 GHz signal from the RF generator. In both cases, the corresponding values measured at zero applied magnetic field were subtracted to remove the contribution arising from direct electromagnetic coupling between the excitation and detection antennas. (d) Circuit level simulation using experimental data - schematic and real and imaginary parts as a function of the 12.22 GHz RF signal phase; Circuit level simulation using: real part of the experimental data for the phase shifts $\Delta\varphi$ of 0$^{\circ}$ and 90$^{\circ}$ (e); imaginary part for the phase shifts $\Delta\varphi$ of 0$^{\circ}$ and -90$^{\circ}$ (f). All measurements and simulations were performed under an applied magnetic field of $\mu_{0}$H = 43 mT.}
\end{figure*}

A second type of experiment was performed using VNA-1 to sweep the frequency between 6 and 15 GHz, while VNA-2 operates as an RF generator (continuous wave) at a fixed frequency of 12.22 GHz. In this configuration, the spin wave generated at 12.22 GHz propagates simultaneously with the broadband spin waves excited by VNA-1 within the same waveguide. The output power for the two VNAs was set at 0.5 mW. The spin wave transmission characteristics for the RF signal at 12.22 GHz on or off is shown in Fig.~\ref{fig3}(a). Fig.~\ref{fig3}(b) and (c) show the real and imaginary parts of the measured data.   
The very good overlap between the traces measured by VNA-1 with and without the 12.22 GHz signal demonstrates yet again that there is no interaction between the spin-waves excited in a linear regime in a broad frequency range with the wave excited at 12.22 GHz. Nevertheless, a strong increase in the spin-wave amplitude at 12.22 GHz is seen as a result of the additional contribution of the power generated by the VNA-2 to the spin-wave excitation.

%The corresponding wavelength of the spin waves excited by VNA-2 at 12.22 GHz for the applied magnetic field is about 2535 nm.Note that the VNA-1 is exciting spin waves with a large spectrum of wavelengths depending on the frequency.
% between the two U-shaped antennas was recorded for different applied magnetic fields, for the RF signal at 12.22 GHz on or off (see
It must be noted that the 12.22 GHz continuous wave signal is not in phase with the swept signal generated by VNA-1, and the result of the interference of the two spin waves depends on the phase difference between them. 

To further analyze this behavior, circuit-level simulations were performed using AWR Microwave Office, based on experimentally measured S-parameters at the external magnetic bias field of 43 mT. The measured S-parameters of the spin-wave device were implemented as a Touchstone block and combined with the 12.22 GHz single-tone signal via an ideal (lossless) power combiner (see Fig.~\ref{fig3}(d)). The single-tone input was also defined as a Touchstone file, incorporating the experimentally measured amplitude and phase at 12.22 GHz, while all other frequency points were treated as open circuits. This approach ensured correct mathematical combination of the respective S-parameter matrices. In addition, a phase shifter was introduced after the single-tone source, allowing controlled variation of its phase prior to combination and enabling extraction of the resulting S-parameters for different relative phase conditions. Fig.~\ref{fig3}(e)-(f) show examples of the real and imaginary parts for different phase shifts $\Delta\varphi$.

%A circuit level simulation was performed using AWR Microwave Office and measured S-parameter data at 43 mT external magnetic biasing field for the Damon-Eshbach configuration. The measured S-parameter results of the spin wave device were introduced as a Touchstone circuit block and combined with the 12.22 GHz single tone through an ideal (lossless) power combiner block. The 12.22 GHz tone was introduced as a Touchstone file with the measured amplitude and phase as it passed through the spin wave device, with all other frequency points set as ideal open circuit. This procedure was needed in order to properly perform the mathematical operations between the two S-parameter matrices. A phase shifter was added after the 12.22 GHz block, permitting the introduction of desired phase shifts before the combiner. This allowed extraction of the combined S-parameter results for a variety of phase shifts of the single tone.

\begin{figure*}
\includegraphics[width=1\textwidth]{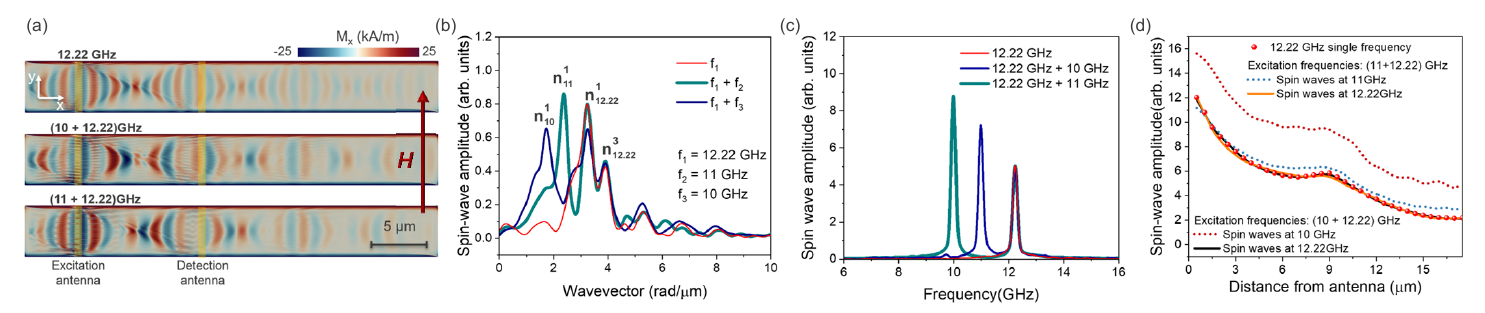}% Here is how to import EPS art
\caption{\label{fig4}(a) Snapshot images of the magnetization oscillation obtained from  micromagnetic simulations of the CoFeB waveguide for the indicated excitations frequencies. (b) Spin-wave amplitude at different frequencies, extracted at the detection antenna shown in (a), located at \qty{10}{\micro\meter} from the excitation. (c) Spin-wave wavevectors in the waveguide at different excitation frequencies, extracted from the micromagnetic simulations. (d) Spin-wave amplitude as a function of propagation distance for the different excitation frequencies. All simulations were performed under an applied magnetic field of $\mu_{0}$H = 43 mT.}
\end{figure*}

The experimental results clearly demonstrate that, in the linear excitation regime, co-propagating spin waves at different frequencies do not exhibit measurable interaction. This behavior is consistent with the superposition principle governing the linearized magnetization dynamics, whereby each frequency component propagates independently within the magnetic waveguide. As a result, multiple spin-wave signals can coexist simultaneously without mutual distortion or cross-talk, even when sharing the same physical channel. 

Nevertheless, for a quantitative in-depth analysis, we performed micromagnetic simulations of co-propagating spin-waves at different frequencies using the MuMax3 framework \cite{MuMax3}. The simulations account for the same geometry and magnetic properties as in the experiments. A mesh size of $9 \times 9 \times 7.5$ $\text{nm}^3$ and an exchange constant $A=18.6$~pJ/m \cite{talmelli2020} were considered. A direct comparison was performed between spin-wave excitation using two simultaneous frequencies (\textit{i.e.}, $f_{1}+f_{2}$,$f_{1}+f_{3}$) and the single-frequency case ($f_{1}$), where $f_{1} =12.22$ GHz, $f_{2} =11$ GHz and $f_{3} =10$ GHz, respectively. The snapshot images of the dynamic magnetization pattern for the different frequencies, shown in Fig.~\ref{fig4}(a), display the typical interference pattern of different propagating modes with different wavevectors. \cite{pirro2011} The wavevectors corresponding for each propagating mode, as extracted from simulations, are shown in Fig.~\ref{fig4}(b). The wavevectors corresponding to the first width mode $n_{1}$ are clearly visible for all three excitation frequencies ($f_{1}$, $f_{2}$ and $f_{3}$), whereas the wavevector associated with the third width mode $n_{3}$ is observed only for $f_{1} =12.22$ GHz. This can be explained by the fact that, for $f_{2}$ and $f_{3}$, the wavevectors of the third width modes start to overlap with $n_{1}$ at 12.22 GHz, making them indistinguishable in the spectral analysis. 

Consistent with the experimental observations, the micromagnetic simulations show that the spin-wave amplitude is not affected by the simultaneous injection of multiple frequencies into the waveguide, as illustrated in Fig.~\ref{fig4}(c). The amplitude of $f_{1}$ modes remains unchanged compared to the case of single-frequency excitation, confirming the absence of interaction in the linear regime. This behavior is maintained over the entire propagation distance along the waveguide (see Fig.~\ref{fig4}(d)), indicating that the coexistence of multiple frequency components does not lead to cumulative effects or degradation of individual modes. As a consequence, the spin-wave signal is not sensitive to the position of the detection antenna, and the transmission characteristics remain consistent regardless of the probing location. \\

In summary, we have demonstrated the linear superposition of independently generated spin waves propagating in a CoFeB waveguide using a fully electrical excitation and detection scheme. Simultaneous excitation by two independent microwave sources produced transmission characteristics that were indistinguishable from those obtained under single-channel operation, confirming that spin waves at different frequencies propagate independently in the linear regime. The experimental observations are supported by theoretical analysis based on the superposition principle and by micromagnetic simulations, which show that the amplitudes and wavevectors of the individual modes remain unchanged during co-propagation. Additional measurements combining broadband frequency sweeps with continuous-wave excitation, together with circuit-level simulations, further clarify the influence of the relative phase between independently generated signals on the measured scattering parameters. The presented results establish a practical framework for the electrical characterization of multiplexed spin-wave systems and provide direct evidence that multiple frequency channels can coexist within a single magnetic waveguide without measurable crosstalk. These findings demonstrate the feasibility of spin-wave frequency-division multiplexing and represent an important step toward parallel magnonic computing, signal-processing, and communication architectures.

\begin{acknowledgments}

The authors would like to acknowledge the support of the European Union Horizon Europe research and innovation program under grant agreement no. 101070417 (SPIDER). A.C.B. also acknowledges the support of the MCID, CNCS/CCCDI - UEFISCDI, project number PN-IV-P8-8.1-PRE-HE-ORG-2023-0042, project number PN-IV-P2-2.1-TE-2023-1239 and the core program project no. PN23070101-2023. L.S. acknowledges the support of RExQTCS CCCDI - UEFISCDI project number PN-IV-P6-6.1-CoEx-2024-0214.
\end{acknowledgments}

\textbf{AUTHOR DECLARATIONS}

\textbf{Conflict of interest}\\
The authors have no conflicts to disclose.

\textbf{DATA AVAILABILITY}\\
The data that support the findings of this study are available from the corresponding author upon reasonable request.

%\nocite{*}
\bibliography{apssamp}% Produces the bibliography via BibTeX.

\end{document}